# Coherent control of optical spin-to-orbital angular momentum conversion in metasurface

Huifang Zhang,[1] Ming Kang,[2,*] Xueqian Zhang,[1,†] Wengao Guo,[3] Changgui Lu,[3] Yanfeng Li,[1] Weili Zhang,[1,4] and Jiaguang Han[1, ‡]

[1]Center for Terahertz Waves and College of Precision Instrument and Optoelectronics Engineering, Tianjin University, Tianjin 300072, China
[2]College of Physics and Materials Science, Tianjin Normal University, Tianjin 300387, China
[3]Advanced Photonics Center, Southeast University, Nanjing 210096, China
[4]School of Electrical and Computer Engineering, Oklahoma State University, Stillwater, Oklahoma 74078, USA

## Abstract:

We propose and experimentally demonstrate that a metasurface consisting of Pancharatnam-Berry phase optical elements can enable the full control of optical spin-to-orbital angular momentum conversion. Our approach relies on the critical interference between the transmission and reflection upon the metasurface to create actively tunable and controllable conversion with a high output via coherent control of the two incident beams. The introduced control methodology is general and could be an important step toward the development of functional optical devices for practical applications.

[*]mingkang@mail.nankai.edu.cn

[†]alearn1988@tju.edu.cn

[‡]jiaghan@tju.edu.cn

***Introduction.***—Spin-orbit interaction(SOI) of light refers to the interplay between the polarization (spin) and spatial (orbital) degrees of freedom of light. Similar to the SOI of electrons occurring in solids[1-3], the interplay between the two distinct angular momentum forms of light reveals many counter-intuitive physical phenomena [4-30].Therefore, understanding and engineering the SOI of light is important to explore the exotic optical responses of materials. Meanwhile, it also provides new opportunities for generating structured states of light[31,32]. Recently, it has been found that the conversion from optical spin to orbital angular momentum can be greatly enhanced by properly designing the inhomogeneous and anisotropic media, which offers a unique route towards controlling the SOI of light in various synthesized materials[10, 11]. The key component of these materials is generally referred to as the Pancharatnam-Berry phase optical element (PBOE), which has been already applied in various architectures, such as gratings, liquid crystals, and metasurfaces[10,11,22-24,30]. Specifically, more recent research highlights the role of the metasurfaces in the realization of SOI, where planar metasurfaces with sub-wavelength thickness display numerous intriguing and unprecedented properties in manipulation of light in a desirable manner[22-24,31,32].

However, one of the challenges in designing a metasurface based on PBOEs is the dilemma between the conversion efficiency and the signal transmittance. In a single-layer metasurface with deep subwavelength thickness, $t \ll \lambda$, huge anisotropic response is required to achieve unitary conversion according to $\sqrt{\varepsilon_y} - \sqrt{\varepsilon_x} = \lambda/(2t)$, which will result in near zero transmittance of the signal due to the increased impedance mismatch, $\lambda/(2t\sqrt{\varepsilon_y}\sqrt{\varepsilon_x}) = \Delta Z$, where $t$ is the effective thickness, $\varepsilon_x$ and $\varepsilon_y$ are the respective permittivities along the two orthogonal axes perpendicular to the propagation direction, and $\lambda$ is the working wavelength of the PBOEs. Thus, it is a great challenge to realize 100% conversion while maintaining a high transmittance. In addition to the afore-mentioned problem, the design also suffers from another inherent problem, where once the structure of the designed metasurface is fixed, active tuning of the conversion processes is normally excluded. As such, efforts are being made to achieve highly efficient spin-to-orbital angular momentum conversions with a high transmittance as well as actively controllable functionality[33-38].

In this context, we report on complete and efficient control over the conversion of optical angular momentum from spin to orbital form in a metasurface system. The design stems from a PBOE reified by a metallic resonator, and if the ohmic loss of the resonator is small enough, it can theoretically support nearly100% conversion and unitary output. Our study shows that conversion efficiency dynamically tunable from 0 to 100% can be obtained by simply varying the relative phase of the two input beams. The presented proposal provides a potential route to actively control the optical spin-to-orbital angular momentum conversion with high performance, and this approach could also be easily extended to other systems, allowing substantial flexibility in the control of the optical spin-to-orbital angular momentum conversion.

*Theoretical model*.—For simplicity, we assume that the investigated metasurface is a free standing planar surface composed of a PBOE array. In each PBOE, the resonant nature is described by a pair of resonant modes without coupling, one oscillating in the *x* direction and the other oscillating in the *y* direction. Such a metasurface can be viewed as an anisotropic metasurface. The optical response of the metasurface to a normally incident plane wave is given by[39]:

$$\mathbf{\Omega q} = \mathbf{Ka}, \mathbf{K}^T\mathbf{q} + \mathbf{Ca} = \mathbf{b}.\quad(1)$$

Here, **a** (**b**) is the complex amplitude of the input (output) field. The resonant complex amplitude of the resonator is described by $\mathbf{q} = (q_x, q_y)$, where the subscripts *x* and *y* indicate the resonator oscillating directions. The response of the resonator is described by:

$$\mathbf{\Omega} = \begin{pmatrix} if_x - if - \gamma_x^s - \gamma_x^d & 0 \\ 0 & if_y - if - \gamma_y^s - \gamma_y^d \end{pmatrix},\quad(2)$$

where $f_j, \gamma_j^s$, and $\gamma_j^d$ (*j*= *x* or *y*) indicate the resonant frequency, scattering loss rate, and dissipation loss rate of the resonator oscillating in the *j* direction. The coupling between the input(output)light and the resonator system is described by $\mathbf{K}\ (\mathbf{K}^T)$, and the direct coupling between the input and output light is illustrated by $\mathbf{C}$. These coupling matrices are given by:

$$\mathbf{K} = \begin{pmatrix} \sqrt{\gamma_x^s} & \sqrt{\gamma_x^s} & 0 & 0 \\ 0 & 0 & \sqrt{\gamma_y^s} & \sqrt{\gamma_y^s} \end{pmatrix},\ \mathbf{C} = \begin{pmatrix} 0 & 1 & 0 & 0 \\ 1 & 0 & 0 & 0 \\ 0 & 0 & 0 & 1 \\ 0 & 0 & 1 & 0 \end{pmatrix}.\quad(3)$$

According to Eq. (1), the relation between the input and output is related by the scattering matrix **S** with $\mathbf{Sa} = \mathbf{b}$. In detail, the scattering matrix **S** is:

$$\mathbf{S} = \mathbf{C} + \mathbf{K}^T\mathbf{\Omega}^{-1}\mathbf{K} = \begin{pmatrix} \mathbf{S}_x & \mathbf{0} \\ \mathbf{0} & \mathbf{S}_y \end{pmatrix} = \begin{pmatrix} r_x & t_x & 0 & 0 \\ t_x & r_x & 0 & 0 \\ 0 & 0 & r_y & t_y \\ 0 & 0 & t_y & r_y \end{pmatrix},\quad(4)$$

where $\mathbf{S}_j$ is the scattering matrix, and $r_j\ (t_j)$ is the reflection(transmission) coefficient in the *j*-polarization channel (*j*= *x* or *y*). Here, it is assumed that $t_j = 1 + r_j$ is satisfied. Due to the absence of coupling, we can separately investigate the response in each polarization channel.

Each scattering matrix $\mathbf{S}_j$ has a symmetric eigenvector $(1,1)$ with eigenvalue $r_j + t_j = 1 + 2r_j$, and an anti-symmetric eigenvector $(1,-1)$ with a constant eigenvalue of -1. If we assume that the dissipation loss rate is very small, $\gamma_j^d \ll \gamma_j^s$, and the probed frequency $f$ is very close to $f_x$ but far away from $f_y$, $|f_x - f| \ll \gamma_x^s \pm \gamma_x^d$ and $|f_y - f| \gg \gamma_y^s \pm \gamma_y^d$, it can be deduced that the symmetric eigenvalue in the $x$-polarization channel has almost a $\pi$ phase difference from that in the $y$-polarization channel, and meanwhile, their output amplitudes are all close to 1. As a result, such a metasurface can serve as an effective half-wave plate under symmetric incidence, i.e., a perfect PBOE with high output. Here, we call it a coherent Pancharatnam-Berry phase optical element(C-PBOE).

Once the basic element of the C-PBOE is designed, the conversion from spin to orbital angular momentum can be engineered by locally rotating each antenna by an angle $\beta(\mathbf{r})$ in the clockwise (counter-clockwise) direction to generate antenna arrays. Like the formulation in the transmission matrix[10,11], the local scattering matrix $\mathbf{S}(\beta(\mathbf{r}))$ can be represented by

$$\mathbf{S}(\beta(\mathbf{r})) = \mathbf{R}^{-1}(\mathbf{C} + \mathbf{K}^T\mathbf{\Omega}^{-1}\mathbf{K})\mathbf{R}, \quad (5)$$

where the rotational matrix $\mathbf{R}$ under scattering approach is

$$\mathbf{R} = \begin{pmatrix} \cos(\beta(\mathbf{r})) & 0 & \sin(\beta(\mathbf{r})) & 0 \\ 0 & \cos(\beta(\mathbf{r})) & 0 & \sin(\beta(\mathbf{r})) \\ -\sin(\beta(\mathbf{r})) & 0 & \cos(\beta(\mathbf{r})) & 0 \\ 0 & -\sin(\beta(\mathbf{r})) & 0 & \cos(\beta(\mathbf{r})) \end{pmatrix}. \quad (6)$$

For example, under a coherent input, $\mathbf{a} = (1, e^{i\psi}, i, ie^{i\psi})$, the output $\mathbf{b}$ in the positive direction is

$$0.5\left[r_x + r_y + (t_x + t_y)e^{i\psi}\right]|+\rangle + 0.5\left[r_x - r_y + (t_x - t_y)e^{i\psi}\right]e^{i2\beta(\mathbf{r})}|-\rangle, \quad (7)$$

where $|\pm\rangle$ indicates the left-(right-) handed circular polarization with the form $(1, \pm i)$. If the

metasurface satisfies the previous requirements, the output field in the positive direction is

$$\left[0.5/\left(\gamma_x^d + \gamma_x^s\right)\right]\left[\left(2\gamma_x^d e^{i\psi} + \gamma_x^s\left(e^{i\psi} - 1\right)\right)|+\rangle - \gamma_x^s\left(e^{i\psi} + 1\right)e^{i2\beta(\mathbf{r})}|-\rangle\right].(8)$$

When a symmetric incidence $(1,1,i,i)$ is applied, the amplitude of the co-polarization component $\gamma_x^d/\left(\gamma_x^d + \gamma_x^s\right)$ is close to 0, while the amplitude of the crossed polarization comment $\gamma_x^s/\left(\gamma_x^d + \gamma_x^s\right)$ is close to 1. When an anti-symmetric incidence $(1,-1,i,-i)$ is applied, the amplitude of the co-polarization $\gamma_x^s/\left(\gamma_x^d + \gamma_x^s\right)$ is close to 1, while the amplitude of the crossed polarization is nearly 0. The conversion efficiency defined by $\eta = \left(I_{crossed} - I_{co}\right)/\left(I_{crossed} + I_{co}\right)$ can be continuously tuned from 0 to 100% by just changing the phase difference $\psi$ of the input beams.

***Design of C-PBOE.***—To experimentally verify the wave manipulation ability based on such a C-PBOE, we consider a metasurface constructed by the above C-PBOEs in the terahertz region. It is worth noting, however, that this proposed approach can also work at other frequencies. Similar to the generation of a Q plate with a vortex profile, here, we design the desired rotating profile $\beta(\mathbf{r})$ in a metasurface with a linear phase gradient, and then measure the co-polarization component in the zeroth diffracted order and the crossed polarization component in the first diffracted order with various input phase differences.

Figure1(a)illustrates the proposed C-PBOE metasurface working around 0.76 THz. The metallic C-PBOE is incarnated by a free standing resonant H-shaped aperture antenna without resting on any substrate. The metasurface is made of 10-μmthick nickel (only 1/40 of the wavelength in free space) with dimensions of $a = 120\,\mu m$, $b = 100\,\mu m$, and $w = 25\,\mu m$. The sample is fabricated using conventional photolithography and metallization process with a schematic of the antenna and the defining parameters shown in the upper row of Fig. 1(a).

The sample is then characterized in a broadband (0.2-2.5 THz) fiber-based terahertz time-domain interferometer. Figure 2 illustrates the experimental setup. A fiber laser source (T-Light, Menlo Systems) produces a 75fsultrashort pulse train centered at 1560 nm. The terahertz pulses are generated and detected by a fiber-based photoconductive transmitter and receiver pair. The generated terahertz wave is first collimated by lens L1 and then split into two beams B1 and B2 by a high resistance silicon wafer SW1, after whichB1 and B2 are adjusted to coincide with each other by two gold-coated mirrors. The sample is placed just in this beam-overlapping region and is normally illuminated by B1 and B2 from opposite sides. By adjusting the position of the samples along the coincided beams with a translation stage, the phase difference between B1 and B2at the sample position can be controlled. Another silicon wafer SW2 is placed at the right side of the sample to guide the transmitted and reflected beams into the receiver. To minimize the amplitude difference, a third silicon wafer(as an attenuator)with a proper tilt angle is placed either in B1 or B2 to attenuate the beam with the larger amplitude. To allow the polarization analysis, five wire-grid linear polarizers, LP1 to LP5 are applied.

Figures 1(b) and (c) show, respectively, the measured amplitude transmissions with horizontal and vertical polarization of the C-PBOE array with a period of $d=200$ μm under normal incidence. The sample exhibits an obvious resonance at $f_x = 0.76$ THz in the $x$-polarization channel, but no resonance in the $y$-polarization channel. The ratio of the transmittances around 0.76 THz between the $x$- and $y$- polarization channels is about 100, indicating that it can serve as a linear polarizer for one beam incidence. Under coherent symmetric illumination in horizontal or vertical polarization channel, as shown in Fig. 1(d), the outputs in the two polarization channels show little difference around the resonance, and maintain a high output amplitude with nearly 100%. However, the output phase difference under coherent symmetric control shows sharp changes due to the resonant nature, see Fig. 1(e). More importantly, the $\pi$ phase difference around 0.76 THz indicates that the proposed antenna can serve as an efficient C-PBOE as discussed above. Because the aperture antenna is the complementary structure to the dipolar strip antenna, we should exchange the roles of reflection and transmission in the fitting process according to the Babinet principle. The fitted analytical results agree well with the experimental results, as shown in Figs.1(b)-(e), where the fitted microscopic parameters are $f_x = 0.7589$ THz, $\gamma_x^s = 0.08571$ THz, $\gamma_x^d = 0.001957$ THz, $f_y = 1.373$ THz, $\gamma_y^s = 0.0851$ THz, and $\gamma_y^d = 0.002$ THz, respectively, satisfying the approximate theoretical requirements at the probed frequency.

*Linear phase gradient.*—Similar to a Q plate, the simple and effective approach to verify the C-PBOE is to generate an optical spin-dependent phase profile under circular polarization incidence. Here, we generate a linear phase gradient metasurface to efficiently bend the output beam at a desired angle. To realize such a linear phase gradient, eight H-shaped aperture antennas with a fixed rotational angle

$$\beta(x) = -x\pi/d, \, x = nd \, (n=1,2,...,7), \tag{9}$$

are designed to form one super cell of the metasurface. The period is $D = 8d = 1.6$ mm along the $x$ direction, and $d = 0.2$ mm along the $y$ direction, as illustrated in Fig. 3(a).

Based on the above discussion, under one beam circular polarization excitation, the

transmitted field in the far-field region can be described by

$$E = \frac{t_x+t_y}{2}\sum_{n=1}^{N}e^{-i\varphi(\mathbf{r},\mathbf{r}')}/\sum_{n=1}^{N}e^{-i\varphi(\mathbf{r},\mathbf{r}'_0)}|\pm\rangle + \frac{t_x-t_y}{2}\sum_{n=1}^{N}e^{-i\varphi(\mathbf{r},\mathbf{r}')\pm 2i\beta(\mathbf{r})}/\sum_{n=1}^{N}e^{-i\varphi(\mathbf{r},\mathbf{r}'_0)}|\mp\rangle, \quad (10)$$

where $\mathbf{r}'$ denotes the far-field position $(x',y',z')$, $\mathbf{r}'_0$ is the far-field position $(0,0,z')$, $\mathbf{r}$ indicates the position of the resonant antenna, $\varphi$ is the accumulated propagation phase from the antenna to the far-field position, $N$ is the total number of the working antennas of the fabricated sample, $|\pm\rangle$ indicates the co-polarization, and $|\mp\rangle$ indicates the crossed polarization. The transmission normalized by the received light without the metasurface at $\mathbf{r}'_0$ contains co-polarization component at the zeroth diffraction order, and crossed polarization component at the first diffraction order. The analytical and experimental results are shown in Figs. 3(b) and (c), respectively. The residual power in the co-polarization channel is due to the fact that a single layer metasurface with deep sub-wavelength thickness is very difficult to realize perfect conversion from optical spin to orbital angular momentum. However, under coherent symmetric incidence $(1,1,\pm i,\pm i)$, the output field in the positive direction in the far-field region can be approximately described by

$$E = \frac{r_x+r_y+t_x+t_y}{2}\sum_{n=1}^{N}e^{-i\varphi(\mathbf{r},\mathbf{r}')}/\sum_{n=1}^{N}e^{-i\varphi(\mathbf{r},\mathbf{r}'_0)}|\pm\rangle$$
$$+\frac{r_x-r_y+t_x-t_y}{2}\sum_{n=1}^{N}e^{-i\varphi(\mathbf{r},\mathbf{r}')\pm 2i\beta(\mathbf{r})}/\sum_{n=1}^{N}e^{-i\varphi(\mathbf{r},\mathbf{r}'_0)}|\mp\rangle \quad .(11)$$

As an example, Figs. 3(d) and (e) illustrate, respectively, the analytical and experimental results under symmetric incidence $(1,1,i,i)$. It is clearly seen that the zeroth order in the co-polarization state is strongly suppressed by the interference. In particular, the zeroth diffraction order power tends to zero around 0.76 THz, implying that no residual power is lost in the co-polarization channel and perfect polarization conversion is achieved. Meanwhile, the first diffraction order intensity in the crossed polarization state around 0.76 THz is strongly boosted due to constructive interference. Such results clearly verify the perfect conversion from optical spin to angular momentum. To clearly show the perfect conversion, we plot the experimentally measured far-field intensities at 0.76 THz as a function of output angle under one beam and two symmetric beam incidences in Figs. 3(f) and (g), respectively. Under one beam excitation, the transmitted power is divided into both the zeroth and first diffraction orders with comparable intensities. In contrast, under symmetric beam excitation, most of the output power is in the first diffraction order with nearly no power in the zero order.

To experimentally obtain the results for the linear-phase-gradient C-PBOE metasurface, we reform the above-mentioned interferometer to a spatial angle-resolved interferometer. The difference is that a rotation stage is introduced so as to allow the collection of the output terahertz beams that are diffracted into different directions, as shown in Fig. 2. The center of the rotation stage is adjusted to coincide with the mirror position of the center of the linear-phase-gradient metasurface with respect to SW2. To achieve the results under circular polarization incidence, we first measure the detailed far-field output, including amplitude and phase under linear polarization base. Then, the received signals are transformed to the circular polarization base, like the transmission matrix transformation from linear to circular polarization base[10]. In Fig. 3, it can be seen that there is slight power distributed in the other first diffracted order in the experimental result, and this is mainly due to the little difference of the incident amplitudes between the horizontal and vertical polarization channels.

In addition to the perfect conversion under symmetric coherent control, the proposed approach could also realize tunable conversion, which is very difficult in conventional metasurfaces. In this case, the output intensity that is divided into the zeroth and first diffraction orders, can be dynamically controlled by varying the phase difference between the two input beams, as shown in Fig.4. When the phase difference increases from 0 to $\pi$, the output intensity in the zeroth diffraction order increases monotonically from zero to unitary, while the intensity in the first diffraction order monotonically decreases from unitary to zero. The experimental results agree well with the analytical predictions. Figure 5 illustrates the conversion efficiency $\eta = (I_R - I_L)/(I_R + I_L)$ as the phase difference $\phi$ changes. It is obvious that the conversion efficiency can be continuously tuned from 0 to 100%. The experimental results are consistent with the theoretical prediction, where the experimental conversion efficiency is estimated by the maximum intensities for the zeroth and first diffracted orders.

In conclusion, we provide an efficient approach to engineer the conversion from optical spin to orbital angular momentum in designed metasurfaces constructed by the C-PBOEs. By varying the relative phase of the two input beams, the conversion efficiency can be continuously tuned from 0 to 100%. As a primary example, a metasurface working in the terahertz region is designed, and a metasurface with linear phase gradient is both theoretically and experimentally demonstrated. Due to the controllable conversion of the C-PBOEs, the system constructed by such elements could find applications in various systems to efficiently control the polarization as well as the intensity and phase distributions of light.

**Acknowledgements**

This work was supported by the National Key Basic Research Program of China (Grant No. 2014CB339800), the National Science Foundation of China (Grant Nos.11304226，61138001, 61107085, 61107053, 61422509, and 61420106006), the Program for Changjiang Scholars and Innovative Research Team in University (Grant No. IRT13033), the Major National Development Project of Scientific Instruments and Equipment (Grant No. 2011YQ150021). H. Z., M. K., and X. Z. contributed equally to this work.

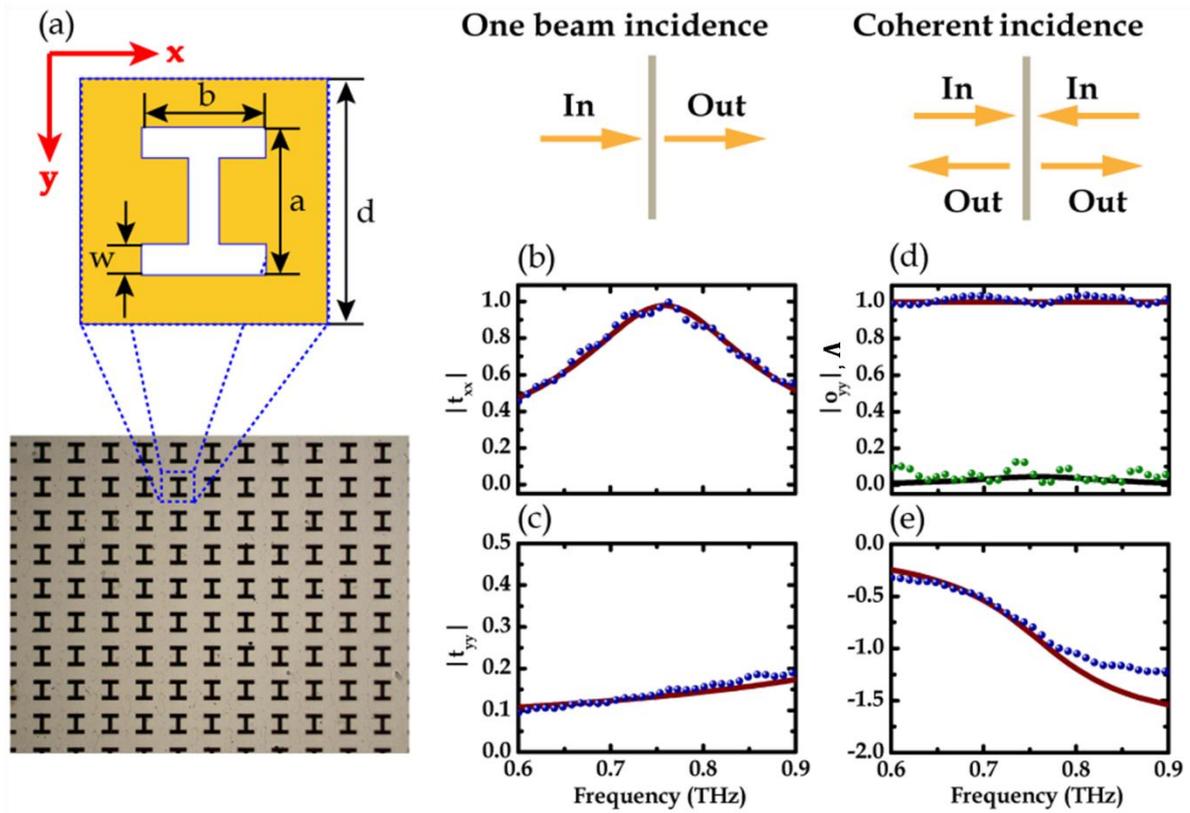

**FIG. 1. Design and characterization of a C-PBOE metasurface.** (a) Microscopic image of the fabricated metasurface with periodic resonant aperture antennas. (b) and (c) Analytical (solid lines) and experimental (scatters) amplitude transmissions in horizontal ($x$) and vertical ($y$) polarization channels under one beam normal incidence. (d) Analytical (solid lines) and experimental (scatters) output amplitudes in vertical polarization channel and the intensity difference of the output between the vertical and horizontal polarization channels under coherent symmetric incidence $\Lambda = \left| |o_{yy}|^2 - |o_{xx}|^2 \right| / \left| |o_{yy}|^2 + |o_{xx}|^2 \right|$. (e) Analytical (solid line) and experimental (scatters) phase difference of the output between the vertical and horizontal polarization channels under coherent symmetric incidence $\left( \mathrm{Arg}(o_{yy}) - \mathrm{Arg}(o_{xx}) \right) / \pi$. Around the working frequency 0.76 THz, the metasurface can serve as an effective half wave plate with high output intensity under coherent symmetric incidence, while it serves as an effective linear polarizer under one beam incidence.

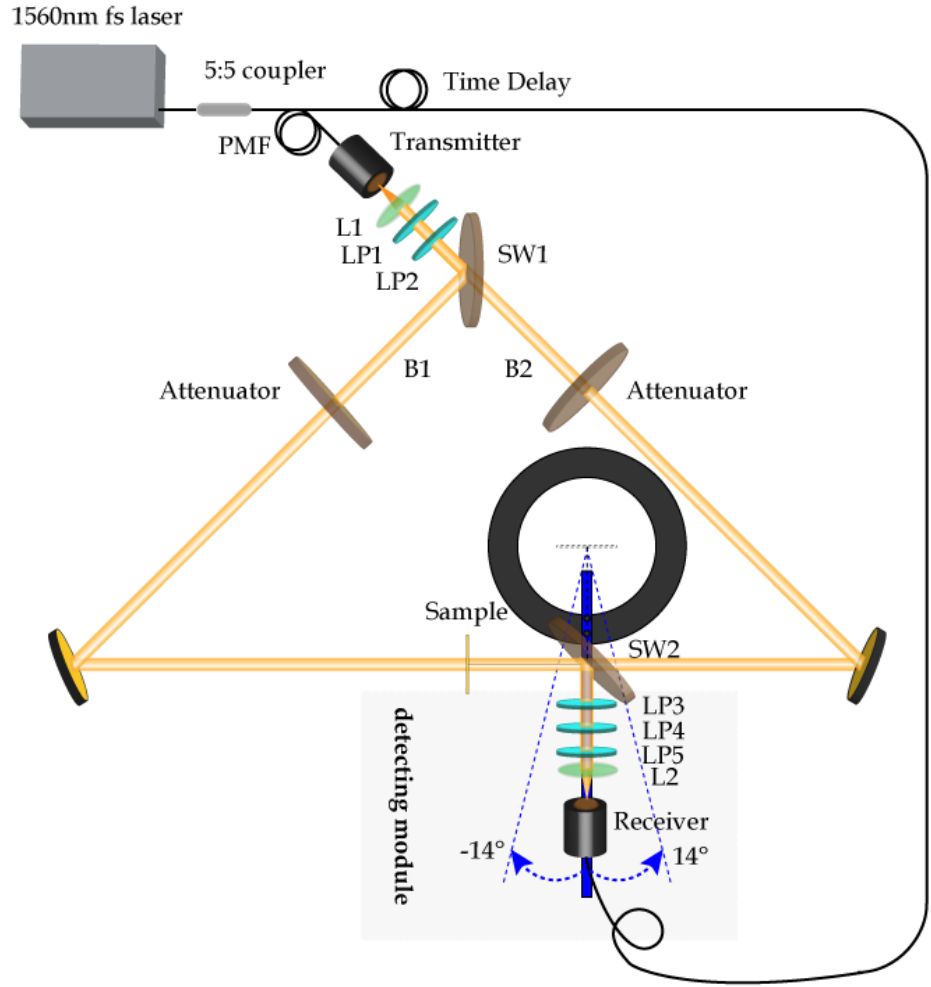

**FIG. 2. Schematic experimental setup for the far-filed measurement of the proposed metasurfaces under coherent control.** Ln (n=1,2) represents terahertz lens, LPn(n=1,2,3,4,5) indicates wire-grid terahertz linear polarizer, and SWn (n=1,2,3) is high resistance silicon wafer to serve as a beam splitter. The incident light from the transmitter is divided into two beams B1 and B2 by SW1. Then, the two beams B1 and B2 are normally incident onto the sample from opposite sides. The detecting module measures the output signal in the far-field. In measuring the metasurface in Fig. 1a, the detecting module just measures the normal output field. However, in measuring the metasurface in Fig. 3a, a rotating stage and a rail are introduced into the detecting module to measure the output field towards different directions. Here, L1 and L2 are combined to change the incident polarization to either *x* or *y* polarization,

while L3-L5 are combined to extract the $x$ and $y$ components of the output polarization.

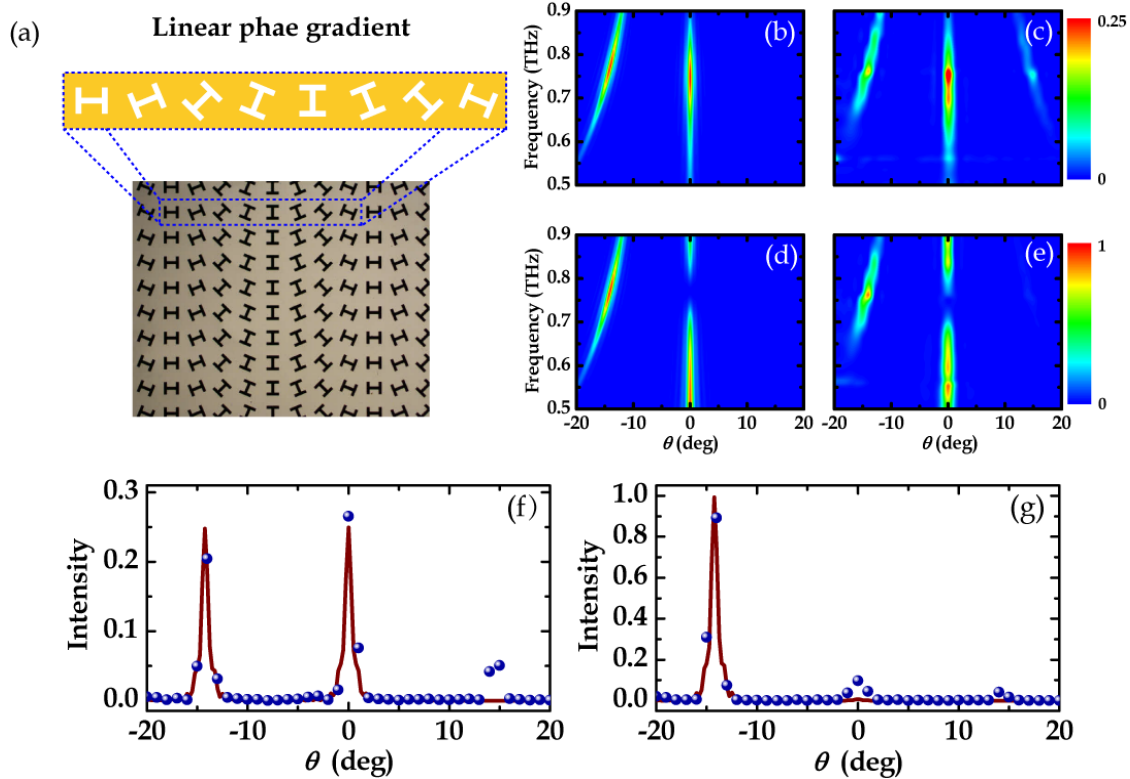

**FIG. 3. Design and characterization of a C-PBOE metasurface with a linear phase gradient.** (a) Microscopic image of the fabricated C-PBOE metasurface with a linear phase gradient. The super cell of the metasurface comprises eight Ni H-shaped aperture antennas with a period of $D = 8d = 1.6$ mm along the $x$ direction and a period of $d = 0.2$ mm along the $y$ direction. Analytical (b, d) and experimental (c, e) far-field intensity profiles of the output beams in the positive direction under one beam $(1,0,i,0)$ and coherent symmetric $(1,1,i,i)$ excitations, respectively. (f) and (g) Analytical (solid line) and experimental (scatters) far-field intensities at 0.76 THz under one beam $(1,0,i,0)$ and coherent symmetric $(1,1,i,i)$ excitations, respectively.

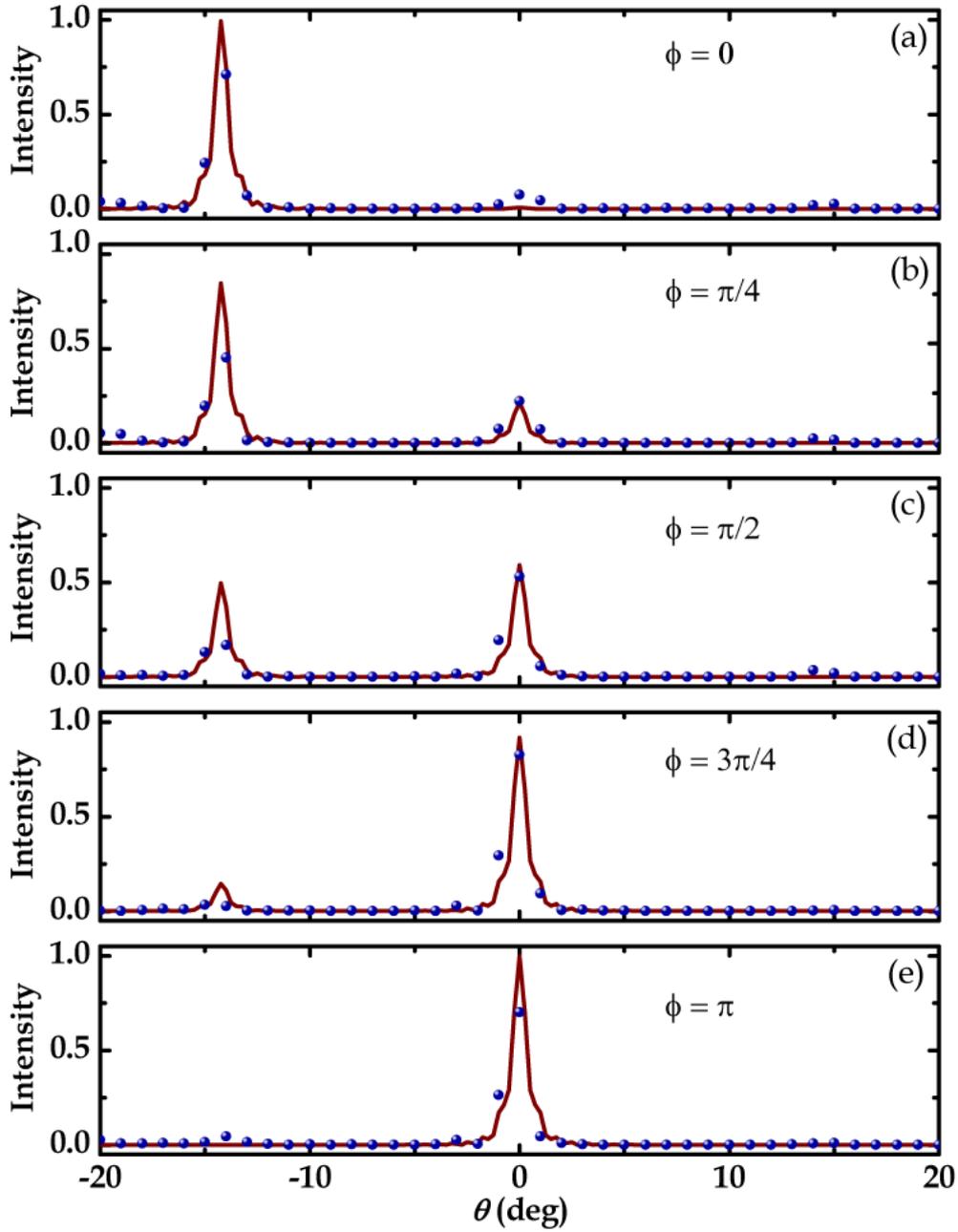

**FIG. 4.The C-PBOE metasurface with a linear phase gradient under phase control.** (a)-(e) Analytical (solid line) and experimental (scatters) far-field intensity profiles of the output fields at 0.76 THz in the positive direction under coherent $(1, e^{i\phi}, i, ie^{i\phi})$ excitation with different phase differences $\phi$.

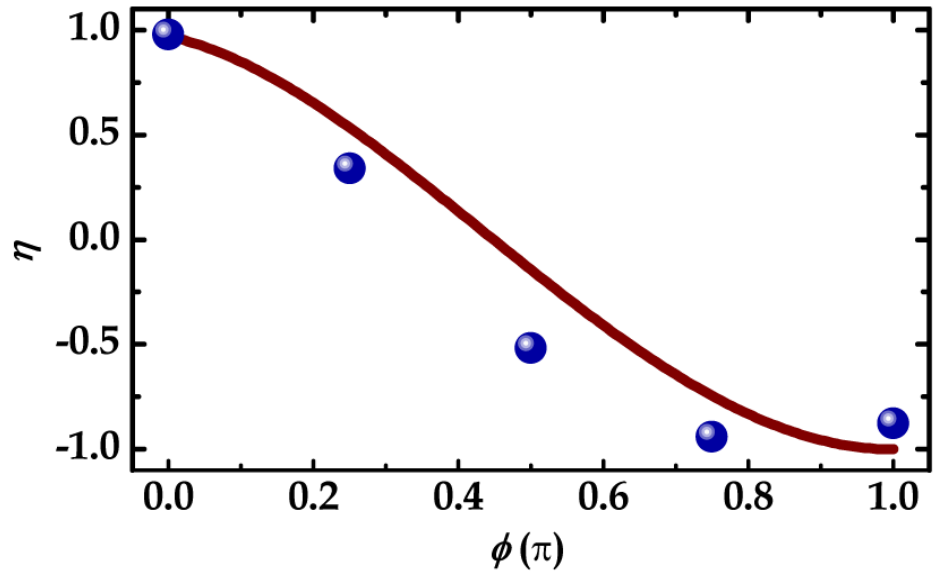

**FIG. 5.** The polarization conversion ratio $\eta = (I_R - I_L)/(I_R + I_L)$ under phase control.